\documentclass[preprint,authoryear,10pt]{elsarticle}

\usepackage{natbib}
\usepackage{epsfig}
\usepackage{graphicx}
\usepackage{bm}
\usepackage{amsmath,amssymb,amsthm,amsfonts,mathrsfs,textcomp,mathtools}
\usepackage{url}
\usepackage{multirow}
\usepackage{subfig}
\usepackage{aas_macros}

\newcommand{\clomon}{{\sc clomon}} % CLOMON in small caps
\usepackage{booktabs}
\usepackage{fullpage}
\usepackage{hyperref}

\journal{Astronomy \& Astrophysics}

\begin{document}

\title{Yarkovsky effect detection and updated impact hazard assessment
  for Near-Earth Asteroid (410777)~2009~FD}

\author[sds,pi]{Alessio Del Vigna}
\ead{delvigna@spacedys.com}

\author[jpl]{Javier Roa}
\author[jpl]{Davide Farnocchia}
\author[esa,inaf]{Marco Micheli}
\author[hawaii]{Dave Tholen}
\author[sds]{Francesca Guerra}
\author[oca]{Federica Spoto}
\author[iaps,ifac]{Giovanni Battista Valsecchi}

\address[sds]{Space Dynamics Services s.r.l., via Mario Giuntini,
  Navacchio di Cascina, Pisa, Italy}

\address[pi]{Dipartimento di Matematica, Universit\`a di Pisa, Largo
  Bruno Pontecorvo 5, Pisa, Italy}

\address[jpl]{Jet Propulsion Laboratory, California Institute of
  Technology, 4800 Oak Grove Drive, Pasadena, 91109 CA, USA}

\address[esa]{ESA NEO Coordination Centre, Largo Galileo Galilei, 1,
  00044 Frascati (RM), Italy}
\address[inaf]{INAF - Osservatorio Astronomico di Roma, Via Frascati,
  33, 00040 Monte Porzio Catone (RM), Italy}

\address[oca]{Universit\'e C\^ote d'Azur, Observatoire de la C\^ote
  d'Azur, CNRS, Laboratoire Lagrange, Boulevard de l'Observatoire,
  Nice, France}

\address[iaps]{IAPS-INAF, via Fosso del Cavaliere 100, 00133 Roma,
  Italy}

\address[ifac]{IFAC-CNR, via Madonna del Piano 10, 50019 Sesto
  Fiorentino, Italy}

\address[hawaii]{Institute for Astronomy, University of Hawaii, 2680
  Woodlawn Drive, Honolulu, HI 96822, USA}

\date{Received xxx; accepted xxx}

\begin{small}
    \begin{abstract}
        Near-Earth Asteroid (410777)~2009~FD is a potentially
        hazardous asteroid with possible (though unlikely) impacts on
        Earth at the end of the 22nd century. The astrometry collected
        during the 2019 apparition provides information on the
        trajectory of (410777) by constraining the Yarkovsky effect,
        which is the main source of uncertainty for future
        predictions, and informing the impact hazard assessment. We
        included the Yarkovsky effect in the force model and estimated
        its magnitude from the fit to the (410777) optical and radar
        astrometric data. We performed the (410777) hazard assessment
        over 200 years by using two independent approaches: the NEODyS
        group adopted a generalisation of the Line Of Variations
        method in a 7-dimensional space, whereas the JPL team resorted
        to the Multilayer Clustered Sampling technique. We obtain a
        4-$\sigma$ detection of the Yarkovsky effect acting on
        (410777), which corresponds to a semimajor axis drift of
        $(3.8 \pm 0.9)\times 10^{-3}$~au/Myr. As for the hazard
        results of both teams, the main impact possibility in 2185 is
        ruled out and the only remaining one is in 2190, but with a
        probability smaller than $10^{-8}$.
    \end{abstract}
\end{small}
\maketitle

\begin{small}{\noindent\bf Keywords}: Impact Monitoring, Line Of
    Variations, Monte Carlo, Virtual impactors, Yarkovsky effect
\end{small}

%
%-------------------------------------------------------------------

\section{Introduction}
Asteroid (410777)~2009~FD is known to have a non-negligible chance of
impacting the Earth in the late 22nd century. With the optical
observations up to 2010 and a purely gravitational model, the
resulting orbital solution was very well-constrained and allowed the
existence of several Virtual Impactors (VIs, \cite{milani:clomon2}),
in particular between 2185 and 2196, with the highest impact
probability attained in 2185.

Adding the optical observations up to April 2014 and the radar Doppler
measurement of Arecibo performed on April 7, 2014, the orbital
uncertainty had shrunk in such a way that the biggest VI would have
been ruled out. However, \cite{spoto:410777} showed that the Yarkovsky
effect was a key source of future uncertainty because of the length of
the impact analysis time interval and the presence of deep planetary
encounters. Unfortunately, the observational data set available in
2014 was not enough to directly detect the Yarkovsky effect from the
orbital fit to the astrometry.

Therefore, to model the Yarkovsky effect \cite{spoto:410777} relied on
the available physical characterisation of (410777) and general
properties of the Near-Earth Asteroid population. When accounting for
the Yarkovsky effect, the 2185 VI could not be ruled out and had an
impact probability $IP = 2.7\times 10^{-3}$ and a Palermo Scale
$PS=-0.43$, the highest in the risk list of both
\clomon-2\footnote{\url{http://newton.spacedys.com/neodys/index.php?pc=4.1}}
and Sentry\footnote{\url{http://cneos.jpl.nasa.gov/sentry/}}. Such a
high value of the Palermo Scale was mainly due to the estimated mass
of the asteroid. Based on WISE data, the diameter and the albedo of
(410777) were estimated as $(472\pm 45)$~m and $(0.010\pm 0.003)$
respectively \citep{mainzer:neowise}, yielding a nominal mass of
$8.3\times 10^{10}$~kg by assuming a density of $1.5$~g/cm$^3$, as in
\cite{spoto:410777}.

The next recomputation of the impact monitoring results by \clomon-2
and Sentry occurred in the early 2016. At that time the astrometric
data set of (410777) included data from two further apparitions: one
in late December 2014 and one in October-December 2015, which also
included radar measurements from Arecibo and Goldstone
\citep{cbet:radar}. These data led to two main improvements.
\begin{itemize}
  \item The 2015 radar observations revealed that (410777) is a binary
    system, with the diameter of the two components roughly 120-180~m
    and 60-120~m \citep{cbet:radar}. These values led to a nominal
    estimate of $3.2\times 10^9$~kg for the mass of the whole system
    and, in turn, to a decrease of the Palermo Scale. Note that the
    updated measurement of the diameter is significantly lower than
    previously estimated from NEOWISE data \citep{mainzer:neowise} and
    adopted by \cite{spoto:410777}. The inaccuracy of the previous
    estimate is explained since the WISE measurements of (410777) only
    correspond to few of the brightest detections and did not cover
    the full rotational curve of (410777)\footnote{Mainzer and
      Masiero, personal communication.}. This under-sampling is known
    to lead to diameter overestimates
    \citep[Figure~4]{mainzer:neowise}.
  \item The constraint on the Yarkovsky effect from the astrometry
    became stronger than that from the physical model (see
    \cite{delvigna:yarko}), which is further complicated by the fact
    that (410777) is a binary \citep{vokro:yarko_binary}. As for the
    impact monitoring results, the set of VIs remained essentially the
    same as before.
\end{itemize}

The current astrometric data set available for (410777) includes 39
additional optical observations from the 2019 apparition. We present
the updated hazard assessment for (410777), achieved independently
with the Line Of Variations (LOV) method, used by NEODyS, and with the
Multilayer Clustered Sampling (MLCS, \cite{roa}) technique, adopted by
JPL.

\section{Astrometry}
\label{sec:astrometry}

Asteroid (410777) was initially discovered by the La Sagra survey in
March 2009, and then linked to precovery observations in the same
apparition by the Spacewatch survey, which is consequently credited as
discovery site. The available observational arc now extends for 10
years and 5 separate oppositions, from discovery to the latest
opposition in 2019. The 2009 and 2014-2015 apparitions are responsible
for the majority of the astrometry, with the latter also containing
all the radar detections.

Most observations collected so far during the 2019 apparition were
reported by the Spacewatch team, or by serendipitous detections by
Near-Earth Objects surveys. In addition, our group provided
observations from ESA's Optical Ground Station (Minor Planet Center
code J04) and the University of Hawaii 2.2 meter telescope (Minor
Planet Center code T12). For these observations the astrometric
uncertainty of each position is available and has been included in the
orbit determination process (see Table~\ref{tab:astrometry2019}). The
formal uncertainties for code T12 include contributions from the
astrometric solution and the centroiding on the target asteroid, added
in quadrature. Uncertainties caused by variations in the sky
transparency, seeing, and tracking of the telescope are much harder to
quantify, but can be empirically estimated as $0.05''$ primarily in
the along-track direction. In general, differential refraction also
contributes to the uncertainty (see \cite{tholenfarn}), though the
zenith distance of 18~deg for both the 2019 March 09 and April 04
observations suggests that the amount is at most a few hundredths of
an arcsecond. We thus conservatively adopted an uncertainty floor of
$0.10''$ for these three observations. For the other observations no
uncertainty information is available and therefore we assumed standard
weights based on the \cite{veres:2017} error model.

A search for precovery observations in the image archives of various
professional telescopes, executed via the CADC SSOIS interface
\citep{gwin:2012}, did not produce any additional detection. The
object was never brighter than $V=24$ at any of the times where
suitable images were being exposed by any of the $\simeq 50$
professional instruments covered by the archive.

\begin{table}[h!]
    \setlength{\tabcolsep}{2.5pt}
    \begin{center}
        \caption{Observational data of the 2019 apparition, from
          observatories J04 and T12, with their formal
          uncertainties. Table columns: calendar date (year, month,
          and day), right ascension $\alpha$ in hours, minutes and
          seconds, declination $\delta$ in degrees, minutes and
          seconds, uncertainty of $\alpha$ and $\delta$, observatory
          code.}
        \label{tab:astrometry2019}
        \vspace{0.3cm}
            \begin{tabular}{cccccc}
                \toprule
                {\bf Date} & {$\alpha$} & {$\delta$} & {$\sigma_\alpha$} & {$\sigma_\delta$} & {\bf Code}\\
                \midrule
                2019-03-06.907914 & 10:14:23.603 & +17 36 15.77 & $0.30''$ & $0.30''$ & J04\\
                2019-03-06.910166 & 10:14:22.988 & +17 36 18.29 & $0.15''$ & $0.15''$ & J04\\
                2019-03-06.912418 & 10:14:22.326 & +17 36 21.61 & $0.15''$ & $0.15''$ & J04\\
                2019-03-09.439509 & 10:02:15.925 & +18 32 08.44 & $0.036''$ & $0.035''$ & T12\\
                2019-03-09.440157 & 10:02:15.723 & +18 32 09.19 & $0.044''$ & $0.044''$ & T12\\
                2019-04-04.264404 & 07:32:22.014 & +23 44 30.00 & $0.031''$ & $0.031''$ & T12\\
                \bottomrule
            \end{tabular}
    \end{center}
\end{table}

\section{Detection of the Yarkovsky effect}
\label{sec:a2}

For this purpose we modelled the Yarkovsky perturbation by using a
transverse acceleration
\[
    \mathbf{a}_t = \frac{A_2}{r^2} \hat{\mathbf{t}},
\]
where $r$ is the heliocentric distance and $A_2$ is the dynamical
parameter to be estimated from the fit \citep{farnocchia:yarko,
  chesley:yarko, delvigna:yarko}. Furthermore, the force model we
adopted includes the gravitational accelerations of the Sun, the eight
planets, and the Moon based on the JPL planetary ephemerides DE431
\citep{de431}, the perturbations of 16 massive main belt bodies and
Pluto, and the Einstein-Infeld-Hoffmann relativistic model
\citep{moyer2003}.

With the weighting scheme described in Section~\ref{sec:astrometry},
the Yarkovsky parameter obtained by the NEODyS group is
$A_2^{\textrm{NEODyS}} = (7.3\pm 1.7)\times 10^{-14}$~au/d$^2$, which
is very consistent with the value
$A_2^{\textrm{JPL}} = (7.4\pm 1.7)\times 10^{-14}$~au/d$^2$ found by
the JPL team. The difference between the two estimates is caused by
the fact that the JPL orbit solution assumes a 1-second time
uncertainty for each ground-based optical observation, whereas the
NEODyS solution does not\footnote{Observations of GPS satellites at
  T12 on the same nights as the observations of (410777) demonstrate
  clock accuracy at a level well within the uncertainty assumed by
  JPL.}. The NEODyS estimate corresponds to a semimajor axis drift
$da/dt = (3.8 \pm 0.9)\times 10^{-3}\text{ au/Myr}$. Thus the current
astrometry provides a 4-$\sigma$ Yarkovsky detection. Note that this
distribution of the $A_2$ parameter is statistically consistent with a
positive Yarkovsky drift: more precisely, the probability of $A_2>0$
computed by using a Gaussian formalism is $99.998\%$. The positive
value of $A_2$ suggests that (410777) is a prograde rotator
\citep{vokr:yark}.

\section{Impact monitoring with the LOV method}
\label{sec:lov}

The inclusion of the Yarkovsky effect in the dynamical model results
in an initial space with 7 dimensions, corresponding to the six
orbital parameters and $A_2$. The LOV definition of
\cite{milani:multsol} can be extended to spaces with dimension greater
than $6$, as was already done in \cite{spoto:410777}, with an
experimental version the OrbFit software. This capability has
undergone testing and it is now included in the operational version
OrbFit 5.0\footnote{\url{http://adams.dm.unipi.it/orbfit/}}.

The LOV definition remains basically the same: the tangent vector to
the LOV at one of its points is the local weak direction, which is the
eigenvector of the $7\times 7$ covariance matrix related to the
largest eigenvalue. Moreover, cases like (410777), for which the
initial confidence region is very small, allow the use of the linear
approximation of the LOV: a particular direction
$\mathbf{w}\in \mathbb{R}^7$ is selected to approximate the LOV with
the straight line passing through the nominal solution $\mathbf{x}^*$
with direction $\mathbf{w}$. In case of a scattering encounter
\citep[Section~4]{spoto:410777} $\mathbf{w}$ is chosen in such a way
that the spread of the corresponding $b$-plane points
\citep{valsecchi:resret} is maximum, so that we can fully capture the
different dynamical evolution of the orbits after the scattering
encounter along the LOV. In particular, this is obtained by first
computing the weak direction on the scattering $b$-plane and then
selecting one of its infinitely many preimages in the initial elements
space through the semilinear formalism \citep{milani:ident2}. We
sampled the LOV over the interval $|\sigma|\leq \sigma_{max}=5$ and
with a generic completeness level of the VI search
$IP^*=1\times 10^{-7}$. This can be achieved with a step-size that is
inversely proportional to the probability density along the LOV,
resulting in a sampling that is denser around the nominal solution and
more sparse towards the LOV tails \citep{delvigna:compl_IM}. To avoid
low resolution in the tail of the distribution we used a maximum value
for the step-size $\Delta\sigma_{max}=0.01$. This setup led to the
computation of 4719 LOV orbits, to propagate with final time in 2250.

\begin{table}[h!]
    \setlength{\tabcolsep}{2.5pt}
    \begin{center}
        \caption{Impact monitoring results of asteroid
          (410777)~2009~FD with a non-gravitational model including
          the Yarkovsky effect. Table columns: calendar date (year,
          month, and day) of the potential impact, $\sigma$ value of
          the VI location along the LOV, distance of the VI trace from
          the centre of the Earth, local stretching at $\sigma$
          \citep{milani:multsol}, probability of Earth impact, and
          Palermo Scale. The width of the $b$-plane confidence region
          is $< 1$~km, thus not reported.}
        \label{tab:410777}
        \vspace{0.3cm}
            \begin{tabular}{cccccc}
                \toprule
                {\bf Date} & {$\sigma$} & {\bf Distance} & {\bf Stretching} & {$IP$} & {$PS$} \\
                           &            & $(R_\oplus)$    & $(R_\oplus)$      &        &        \\
                \midrule
                2190-03-30.08 & $-4.807$ & $0.57$ & $1.22\times 10^3$ & $5.56\times 10^{-9}$ & $-7.25$\\
                \bottomrule
            \end{tabular}
    \end{center}
\end{table}

The impact monitoring results for (410777) are shown in
Table~\ref{tab:410777}. As anticipated, the 2185 impact possibility
has disappeared whereas the 2190 VI still remains, although located in
the LOV tails and thus with a low impact probability. Indeed, the
inclusion of the 2019 astrometry and the consequent improvement of the
Yarkovsky effect estimate decreased the extent of the LOV projection
on the 2185 $b$-plane and excluded any impact possibility within
$\sigma_{max}=5$. Actually the VI still exists, but located at
$\sigma\simeq 7$, and the impact probability is negligible. Typically,
the effect of a close approach is to separate nearby orbits, thus
increasing the uncertainty at subsequent encounters. The increased
post-2185 uncertainty allows the existence of a VI in 2190.

\section{Impact monitoring with the MLCS technique}
\label{sec:mc}

\begin{figure}
\centering
\includegraphics[width=0.7\linewidth]{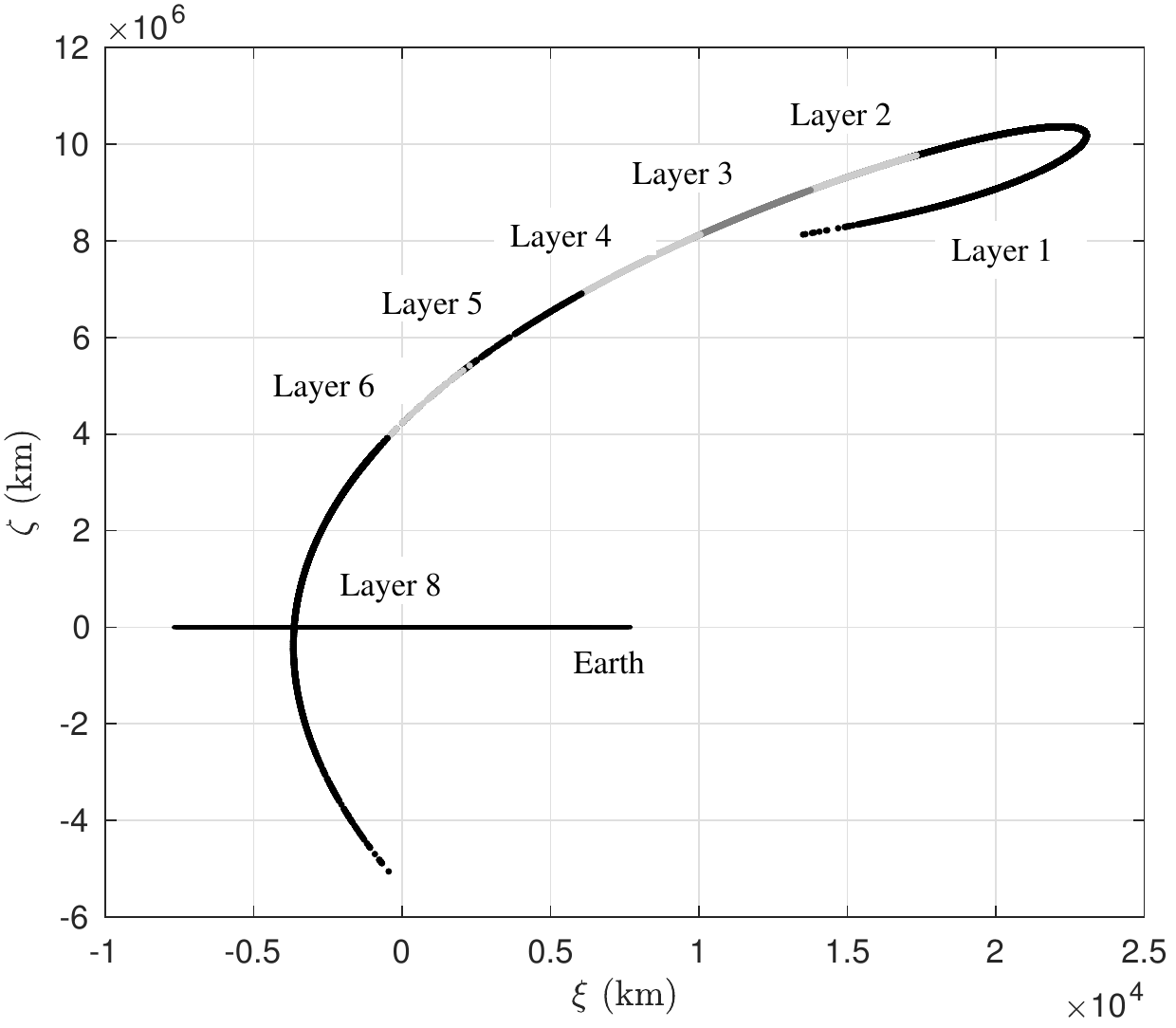}
\caption{Distribution of virtual asteroids on the 2190 $b$-plane
  organised in layers of increasing density. Each set of samples
  satisfies tighter constraints on the value of $A_2$ according to the
  MLCS procedure. Note that the axes use different scales. Layer 7 was
  omitted for clarity.}\label{Fig:MLCS}
\end{figure}

The JPL impact monitoring analysis was performed by using the
Multilayer Clustered Sampling technique \citep{roa}, which is an
efficient alternative to direct Monte Carlo methods. Initially, MLCS
generated the first layer of virtual asteroids by randomly drawing
100~000 samples from the 7-dimensional normal distribution of the
orbital elements and $A_2$. Identifying the 20th percentile of samples
sorted by closest approach distance provides an interval of $A_2$ that
contains the values leading to a close approach in 2190. Next, we
sampled a second layer containing twice as many points as the previous
one, we selected the virtual asteroids for which $A_2$ falls within
the interval defined in the preceding step, and we propagated them to
the 2190 encounter. The 20th percentile of this new set produces a
reduced $A_2$ interval that can be used for further filtering. The
process is repeated sequentially until less than a fraction of
$10^{-5}$ samples satisfies the condition on $A_2$.\footnote{This
  threshold is small enough to efficiently reduce the number of
  virtual asteroids to propagate while ensuring that enough samples
  are left to produce a statistically significant result, since this
  threshold is three orders of magnitude greater than the expected
  impact probability.}  In this case, eight layers were required for
convergence. The final step consists in sampling $10^9$ virtual
asteroids and propagating only those that satisfy the $A_2$ constraint
imposed by the last layer. Figure~\ref{Fig:MLCS} shows how MLCS
iteratively converges to a range of $A_2$ that allows an efficient
exploration of the final layer. As MLCS advances to the next layer,
the samples get closer to the Earth when mapped to the 2190 $b$-plane.

The distribution of samples in each layer is statistically consistent
with the original 7-dimensional distribution in orbital elements and
$A_2$, and the impact probability and its standard deviation can be
computed like in the regular Monte Carlo method. We obtained
$IP=(1.5\pm0.4)\times10^{-8}$. The difference in the result relative
to the NEODyS one is fully explained by the fact that the JPL orbit
solution (JPL~100) assumes a 1-second time uncertainty for optical
observations, which is not part of the NEODyS orbit determination
process. The values of $A_2$ compatible with impact trajectories are
$A_2=(-6.5\pm0.4)\times10^{-15}$~au/d$^2$, suggesting that an impact
is only possible if (410777) were a retrograde rotator.

\section{Estimate of the 2190 keyhole width and location}
\label{sec:opik}

In case there is a pair of resonant returns, a keyhole is one of the
preimages of the Earth impact cross section
\citep{chodas:keyholes}. In other words, if an asteroid passes through
a keyhole during the first encounter, it will hit the Earth at the
subsequent encounter. In our case, we can estimate the width of the
keyhole on the 2185 $b$-plane for impacts in 2190 by using the
analytical theory of close encounters, as in
\cite{valsecchi:resret}. To this end we first compute the length of
the chord $\mathcal{C}$ obtained by intersecting the LOV trace on the
2190 $b$-plane with the Earth impact cross section. Then the value of
the keyhole's width is estimated through the quantity
$\partial\zeta''/\partial\zeta$, where $(\xi,\zeta)$ and
$(\xi'',\zeta'')$ are the coordinates on the two pre-encounter
$b$-planes, respectively. Indeed, this derivative can be seen as the
factor by which the stretching increases in the time span between the
first encounter and the second encounter.

On the 2190 $b$-plane, the chord $\mathcal{C}$ is contained in the LOV
portion between the virtual asteroids bracketing the VI. Let
$(\xi_1'',\zeta_1'')$ and $(\xi_2'',\zeta_2'')$ be their
coordinates. The length of $\mathcal{C}$ is
\[
    \ell_\mathcal{C} = 2\sqrt{b_\oplus^2 - \frac{(\xi_1''\zeta_2'' -
        \zeta_1''\xi_2'')^2}{(\xi_2''-\xi_1'')^2 +
        (\zeta_2''-\zeta_1'')^2}}=2.156 R_\oplus.
\]
The derivative $\partial\zeta''/\partial\zeta$ depends on the
semimajor axis $a$ and the unperturbed geocentric velocity $U$ of
(410777) at the 2185 encounter. In our computations, we assumed
$a=1.1636$~au and $U=16.2$~km/s, which are the values corresponding to
the virtual asteroid closest to the VI. The 2190 $b$-plane is
accessible from the 2185 $b$-plane through the $4:5$ mean motion
resonance between the asteroid and the Earth. Therefore, after $h=4$
revolutions of the asteroid and $k=5$ revolutions of the Earth, a
second close approach takes place. The $4:5$ resonance corresponds to
the post-encounter semimajor axis
$a' = \sqrt[3]{{k^2}/{h^2}} =1.1604\text{ au}$. By the equations in
\cite[Section~4.3]{valsecchi:resret} we obtain
$\partial\zeta''/\partial\zeta \simeq 14.6$, and so the width of the
keyhole is
\[
    w_k \simeq \frac{\ell_{\mathcal{C}}}{\partial\zeta''/\partial\zeta}
    = 943\text{ km}.
\]
The centre of the keyhole, being the preimage of the midpoint of
$\mathcal{C}$, is located at $\zeta_k = -1.333461\times 10^6$~km. This
point belongs to the LOV portion between the two considered virtual
asteroids on the 2185 $b$-plane, which are in turn located close to
the resonant circle corresponding to the $4:5$ resonance\footnote{The
  \emph{locus} of points on a given $b$-plane corresponding to a
  certain mean motion resonance is a circle, with centre on the
  $\zeta$-axis, as shown in \cite{valsecchi:resret}.}.

\begin{figure}[t]
    \centering
    \includegraphics[width=0.7\columnwidth]{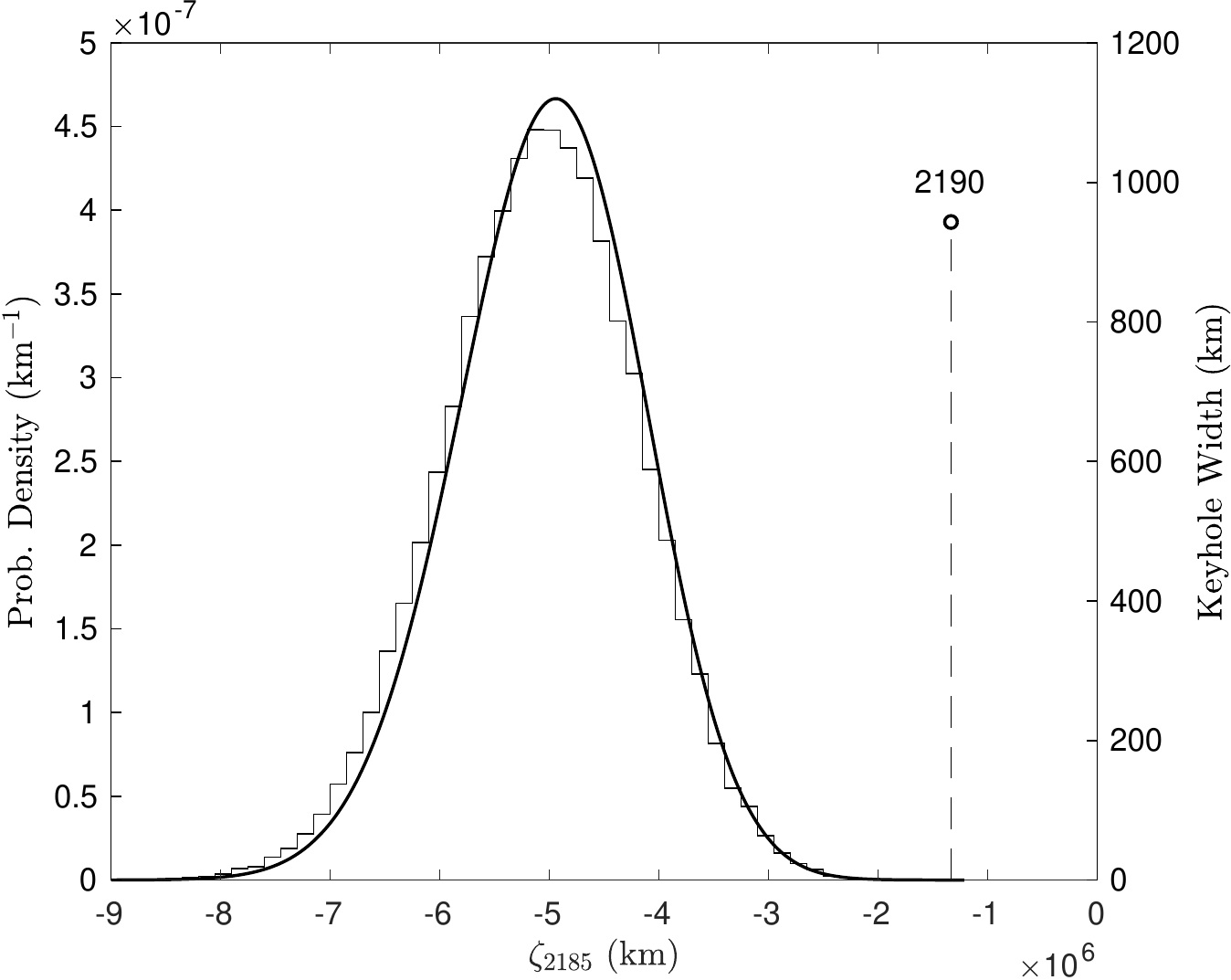}
    \caption{The bold curve is the graph of the NEODyS non-linear
      probability density function $p(\zeta)$ on the 2185 $b$-plane,
      whose scale is reported on the left vertical axis. The 2190
      keyhole is represented by the vertical line segment, which is
      located at the keyhole centre
      $\zeta_k = -1.333461\times 10^6$~km and has height equal to the
      width $w_k=943$~km, with scale on the right vertical axis. The
      histogram represents the distribution based on the virtual
      asteroids of the first MLCS layer from JPL solution 100.}
    \label{fig:pdf}%
\end{figure}

Figure~\ref{fig:pdf} shows the probability density function $p(\zeta)$
on the 2185 $b$-plane, along with the location and width of the
keyhole for impacts in 2190. The position of the keyhole centre on the
tail of the probability distribution implies that the associated
impact probability is very small. In particular, the impact
probability can be estimated as
$IP\simeq w_k p(\zeta_k) \simeq 5.3\times 10^{-9}$, well in agreement
with the result of Table~\ref{tab:410777}.

\section{Conclusions}

We presented the new impact monitoring results for asteroid (410777)
computed by both the NEODyS and JPL teams after the 2019
apparition. These 2019 astrometric observations extended the
observational arc by four years and allowed a 4-$\sigma$ detection of
the Yarkovsky effect through an orbital fit to the astrometry.

To perform the hazard assessment we resorted to two independent
approaches, namely the LOV method for NEODyS and the MLCS technique
for JPL. Both systems ended up with the removal of the 2185 VI, which
was the largest one until the inclusion of the 2019 data. The only
remaining VI is the one in 2190 but, since it lies towards the end of
the LOV ($|\sigma|\simeq 4.8$), it has an impact probability
$< 10^{-8}$, which effectively rules out the corresponding impact.

\section*{Acknowledgements}
A.~Del Vigna and F.~Guerra acknowledge support by the company
SpaceDyS.
Part of this research was conducted at the Jet Propulsion
Laboratory, California Institute of Technology, under a contract
with NASA. Dave Tholen acknowledge NASA support for his
observational work, under the grant No. NNX13AI64G.
The work of F.~Spoto is supported by the CNES fellowship research
programme.
This research has made use of data and/or services provided by the
International Astronomical Union's Minor Planet Center.

%--------------%
% BIBLIOGRAPHY %
%--------------%
\bibliographystyle{elsarticle-harv}
\bibliography{410777_IM}

\end{document}